# Topological solitons in coupled Su-Schrieffer-Heeger waveguide arrays

KHALIL SABOUR[1,*] AND YAROSLAV V. KARTASHOV[2]

[1]*Moscow Institute of Physics and Technology, Institutsky lane 9, Dolgoprudny, Moscow region, 141700, Russia*
[2]*Institute of Spectroscopy, Russian Academy of Sciences, 108840, Troitsk, Moscow, Russia*
*Corresponding author: Khalilsabor1998@gmail.com*

**We investigate the formation of multipole topological solitons at the edges of two and three coupled parallel Su-Schrieffer-Heeger (SSH) waveguide arrays. We show that independent variations of waveguide spacing in the unit cells (dimers) in coupled waveguide arrays result in the emergence at their edges of several topological edge states with different internal symmetry. The number of emerging edge states is determined by how many arrays are in topologically nontrivial phase. In the presence of nonlinearity, such edge states give rise to families of multipole topological edge solitons with distinct stability properties. Our results illustrate that coupling between quasi-one-dimensional topological structures substantially enriches the variety of stable topological edge solitons existing in them.**

Topological insulators are unique materials that behave as insulators in their interior, but at the same time allow currents at their surfaces or edges that are protected by the material's topological characteristics, that grants such currents exceptional robustness against impurities and physical imperfections. These propagation properties have been observed for electromagnetic edge states in various photonic realizations of topological insulators [1-3]. The unique advantage of photonic topological systems is that many of them can possess strong nonlinear response offering powerful tools for control of topological edge states. Particular attention has been paid to self-action effects in topological materials [4] that may lead to the formation of topological edge solitons – hybrid states, whose localization and stability properties can be controlled by their power, that bifurcate from linear edge states in topological bandgap and inherit from them topological protection. For instance, edge solitons have been proposed [5-9] and observed [10-16] in systems based on helical waveguide arrays, truncated optically induced lattices, and higher-order topological structures.

Among the models that can exhibit topological phases, particular attention was devoted to the Su-Schrieffer-Heeger (SSH) lattice that was initially proposed to explain the appearance of localized excitations in polyacetylene [17]. Such structures can be easily realized in photonics using, e.g. waveguide arrays. Nonlinear effects and soliton-like states in SSH lattices have been studied both theoretically [18-20] and experimentally [21-24], so the interplay between nonlinearity and topology in single one-dimensional SSH structures is well understood. At the same time, it was recently shown that coupling between several SSH lattices may substantially enrich the spectrum of the system and dynamics of light propagation in it [25-29], because it qualitatively affects topological properties of the system. Usually such coupling is considered as occurring via one or several lattice sites, as it happens in cross-linked [28] and star-shaped [29] lattices, or two aligned structures placed in close proximity to each other [27]. At the same time, topological soliton formation has never been considered in several parallel SSH lattices, where multiple lattice sites from both structures can interact leading to "global" rather than "local" coupling. In addition to transition from one- to two-dimensional geometries, such systems allow study of localization of light in intriguing situations, where some of coupled lattices are in topological phase, while others are topologically trivial.

In this Letter, we report on topological solitons in parallel coupled dimerized SSH waveguide arrays with focusing nonlinearity. The transition from a trivial to a topological phase in each of these arrays is controlled by independent shifts of the waveguides in dimers that form the array. We consider configurations with two and three SSH arrays and show that their coupling significantly enriches linear spectrum of the system and results in emergence, in the topological phase, of the edge states with different internal symmetry. In nonlinear regime these states give rise to multipole edge solitons, all of which can be stable in proper parameter domains.

We describe the propagation of light in coupled SSH arrays in the material with cubic nonlinearity using the dimensionless nonlinear Schrödinger equation for the light field amplitude $\psi$:

$$i\frac{\partial \psi}{\partial z} = -\frac{1}{2}\left(\frac{\partial^2}{\partial x^2}+\frac{\partial^2}{\partial y^2}\right)\psi - \mathcal{R}(x,y)\psi - |\psi|^2\psi. \quad (1)$$

Here, $x,y$ are the transverse coordinates normalized to the characteristic scale $r_0 = 10\,\mu\text{m}$, $z$ is the propagation distance normalized to the diffraction length $\kappa r_0^2 \approx 1.14\,\text{mm}$, $\kappa = 2\pi n/\lambda$ is the wavenumber in the medium with the background refractive index $n$ (for fused silica $n \approx 1.45$), and $\lambda = 800\,\text{nm}$ is the wavelength. Field amplitude $\psi$ is normalized such that real intensity equals to $I = n|\psi|^2/\kappa^2 r_0^2 n_2$, where $n_2$ is the nonlinear refractive index of the material. Refractive index distribution in several coupled dimerized arrays is described by the function $\mathcal{R}(x,y) = p\sum_{k=1}^{K}\sum_{n=1}^{N}[\mathcal{Q}(x-x_{1n,k},y-y_k)+\mathcal{Q}(x-x_{2n,k},y-y_k)]$, where $K=2,3$ is the number of parallel arrays, $N=9$ is the number of dimers (pairs of waveguides) in each array, $\mathcal{Q}(x,y)=e^{-(x^2+y^2)/a^2}$ are the identical Gaussian functions of width $a=0.5$ (corresponding to $5\,\mu\text{m}$), $x_{1n,k}=-d_\text{x}/2-s_k+2nd_\text{x}$, $x_{2n,k}=+d_\text{x}/2+s_k+2nd_\text{x}$ are the $x$-coordinates of two waveguides in each dimer, $s_k$ is the opposite $x$-shift of the waveguides in $k^{\text{th}}$ array, $2d_\text{x}=6$ is the width of the unit cell (equal in all arrays), $y_k=(2k-K-1)d_\text{y}/2$ is the vertical coordinate, $d_\text{y}=3$ is the vertical array separation, $p=\kappa^2 r_0^2 \delta n/n=5$ is the waveguide depth that corresponds to the refractive index contrast $\delta n \approx 5.5\times 10^{-4}$.

To understand the mechanism of formation of multipole topological edge solitons it is instructive to consider first its linear spectrum. Omitting nonlinear term in Eq. (1) we calculate linear eigenmodes using the ansatz $\psi(x,y,z)=w(x,y)e^{ibz}$, where $b$ is the propagation constant and $w(x,y)$ is a real function, that leads to linear eigenproblem $bw=(1/2)(\partial^2 w/\partial x^2 + \partial^2 w/\partial y^2)+\mathcal{R}w$.

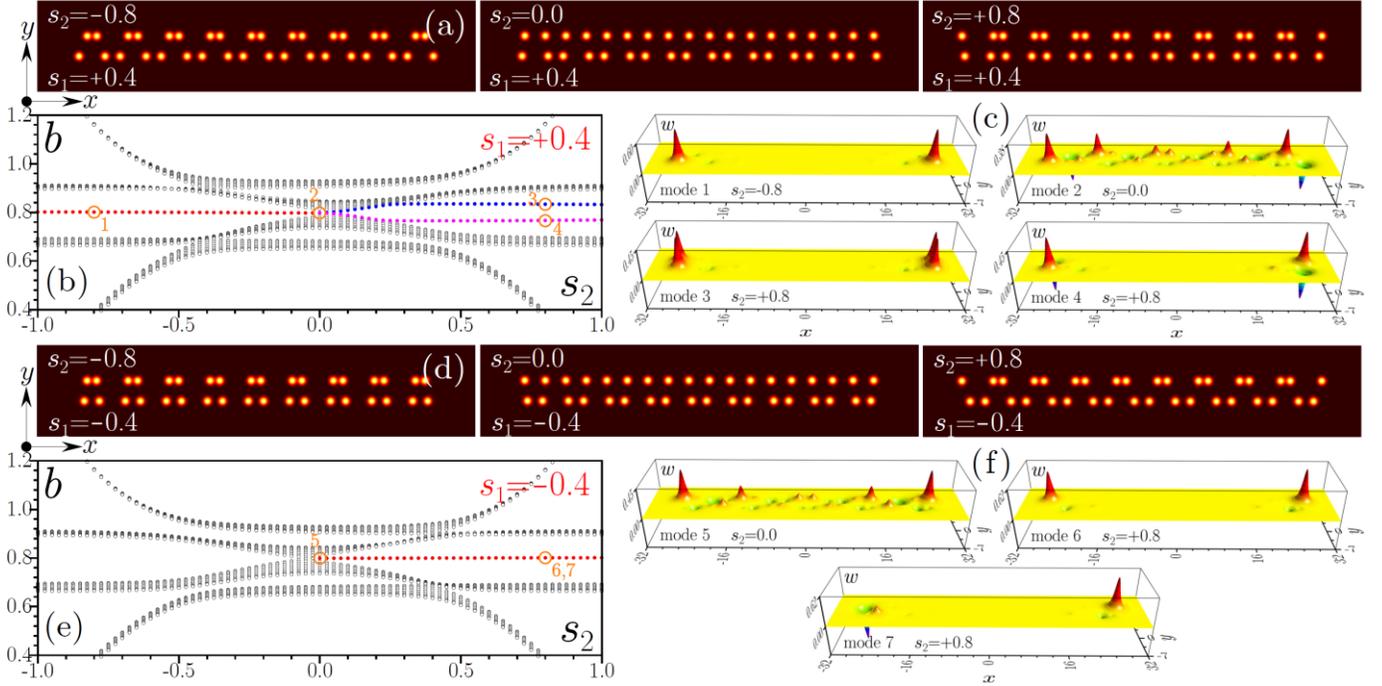

Fig. 1. Two coupled SSH arrays with $s_1=+0.4$ (a) and $s_1=-0.4$ (b) for different $s_2$ values. (b) Propagation constants $b$ of all eigenmodes of the array with $s_1=+0.4$ vs shift $s_2$ and (c) eigenmodes corresponding to orange circles in (b). (e) and (f) show propagation constants $b$ vs shift $s_2$ and representative eigenmodes of the array with $s_1=-0.4$. In (b) and (e) open (gray) dots correspond to bulk states, while red, blue, and magenta dots correspond to edge states.

Examples of arrays, their linear spectra and eigenmodes for the case of $K=2$ arrays are presented in Fig. 1. For each array in this structure, the transition from trivial to topological phase is achieved due to shift of two waveguides in each unit cell (dimer). When this shift is negative, $s_k<0$, the waveguides approach each other and intra-cell coupling between waveguides inside the dimer becomes stronger than the inter-cell coupling between nearest waveguides from neighboring dimers. This is topologically trivial phase [17]. For $s_k>0$, the inter-cell coupling becomes stronger that the intra-cell one, indicating on topological phase [17]. Figures 1(a),(d) illustrate various array configurations when lower array is in topological, $s_1>0$, and nontopological, $s_1<0$, phase, respectively. Because we employ independent shifts $s_k$ in two arrays, further we show the dependencies of propagation constants of all eigenmodes on shift $s_2$ for the fixed shift $s_1$. One can see from Fig. 1(b),(e) that coupling between two arrays substantially enriches spectrum in comparison with single SSH structure [17]. In spectra shown here open (gray) dots correspond to bulk modes, while colored dots correspond to the edge states. When lower array is topological, for $s_2<0$ one observes only one pair (states in such pairs are characterized by in-phase or out-of-phase peaks at the *opposite* array edges, so further we do not distinguish them) of degenerate edge states in the spectrum [see mode 1 in Fig. 1(c)]. Even though upper array is nontopological, this state is well-localized and predominantly concentrated in lower array. At $s_2=0$ the gap closes and all modes become extended [see mode 2 in Fig. 1(c)]. Particularly interesting situation arises at $s_2>0$, when coupling between two arrays results in appearance of two pairs of edge states with in-phase [mode 3 in Fig. 1(c)] and out-of-phase peaks [mode 4 in Fig. 1(c)] in *neighboring* arrays with notably different $b$ values (this difference increases with decrease of spacing $d_\text{y}$). Thus, coupling between topological arrays leads to rich nomenclature of multipole states with different symmetry along the $y$ axis. The total number of topological modes in the spectrum coincides with the number of arrays in topological phase. Notice that light localization degree in $k^{\text{th}}$ array is defined by corresponding shift $s_k$ (it increases with increase of $s_k$) and may be substantially different for different $s_1$ and $s_2$. When lower array is nontopological, $s_1<0$, edge states appear only at $s_2>0$ [Fig. 1(e)] and they are concentrated predominantly in the upper array [see modes 6,7 in Fig. 1(f)].

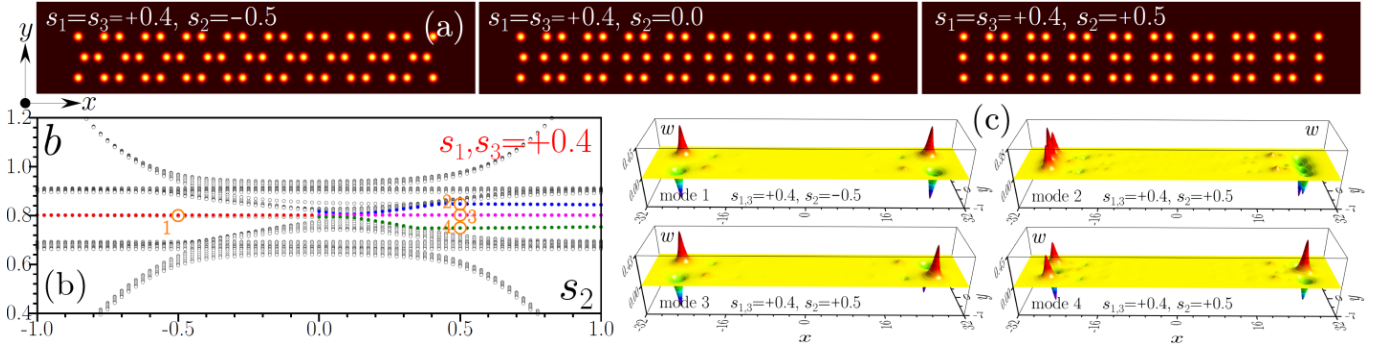

Fig. 2. (a) Examples of three coupled SSH arrays with different shifts of waveguides. (b) Propagation constants of all eigenmodes of the array with $s_1, s_3 = +0.4$ vs shift $s_2$ and (c) eigenmodes corresponding to orange circles in (b). Open dots correspond to bulk states, while red, blue, magenta, and olive dots to edge states.

The spectrum further enriches for $K=3$ arrays. The examples of such structures are presented in Fig. 2(a). The representative linear spectrum in Fig. 2(b) was calculated by varying shift $s_2$ in the middle array, for equal waveguide shifts $s_{1,3} > 0$ in the lower and upper arrays (corresponding to topological phases). At $s_2 < 0$ there are *four* nearly degenerate edge states in topological gap [see representative mode 1 in Fig. 2(c)], with nearly all light concentrated in top and bottom arrays (this explains degeneracy, because these two arrays practically do not feel each other in this regime). At $s_2 > 0$ coupling between three arrays results in the appearance in topological gap of three types of the edge states [modes 2-4 in Fig. 2(c)] with distinct symmetries, from mode 2 with in-phase peaks in all arrays, to mode 4 with out-of-phase peaks. In mode 3 there is no light in middle array. Richer modal structure is a direct consequence of the increase of the number of arrays in the structure. By further increasing number of rows one may expect the transition to fully two-dimensional geometry with extent of the modes in the $y$-direction depending on $K$.

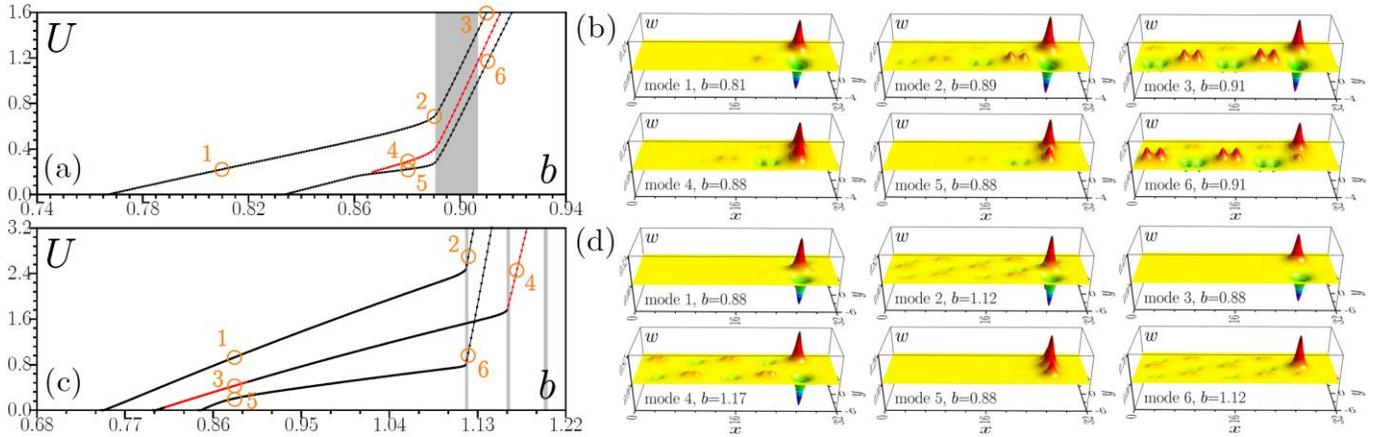

Fig. 3. $U(b)$ dependencies illustrating families of edge solitons in two coupled SSH arrays with $s_1 = +0.4$, $s_2 = +0.8$ (a) and (b) examples of soliton profiles corresponding to the dots in (a). $U(b)$ dependencies for solitons in three coupled SSH arrays with $s_1, s_2, s_3 = +0.8$ (c) and (d) examples of their profiles. Grey regions in (a), (c) correspond to the bands of bulk states. Black branches are stable, while red ones are unstable.

To obtain topological gap solitons, we take into account focusing cubic nonlinearity of the material. Topological solitons bifurcate from linear edge states and remain localized as long as $b$ remains in the topological gap. To characterize soliton families we plot power $U = \iint |\psi|^2 \, dxdy$ vs propagation constant $b$ in Fig. 3(a) for $K=2$ and in Fig. 3(c) for $K=3$ (all arrays are in topological phase in these plots). We also checked stability of solitons by slightly perturbing them at $z=0$ and letting them propagate over very long distance to detect all possible instabilities. Stable edge solitons show propagation with only minor oscillations of amplitude and maintain their structure, while unstable solitons exhibit significant volatility, strong reshaping and considerable amplitude variations. Stable (unstable) soliton branches in Fig. 3 are shown black (red). Notice that in all cases, topological solitons are stable very close to the bifurcation point from linear edge states.

In $K=2$ structure the dipole edge soliton with out-of-phase spots in two arrays [mode 1, Fig. 3(b)] turns out to be stable in the entire topological gap. Here $s_1 < s_2$, so when this soliton enters into the band, it extends mainly in lower array, remaining localized in upper array [mode 2, Fig. 3(b)]. This state remains stable even after crossing of the allowed band. In contrast, its counterpart with in-phase peaks is stable only near the bifurcation point from the linear edge state. Above certain power $U$ strongly asymmetric stable state with dominating peak in upper array [mode 5, Fig. 3(b)] bifurcates downwards from "symmetric" one [mode 4, Fig. 3(b)] that does not change its slope $dU/db$ in the bifurcation point. This is classical symmetry-breaking bifurcation under the action of focusing nonlinearity, akin to the one observed for two waveguides, when asymmetric family that can exist only above certain critical power *smoothly* splits from symmetric one, while its asymmetry gradually increases away from the bifurcation point. Such splitting is accompanied by destabilization of symmetric mode. Asymmetric soliton remains stable even in the band [mode 6, Fig. 3(b)].

In $K=3$ structure at $s_{1,2,3}>0$ one observes three different types of edge solitons [see Fig. 3(c)]. The tripole edge soliton with field changing its sign between neighboring arrays is again stable in the gap and even when it enters into the band [modes 1,2, Fig. 3(d)]. Its dipole counterpart with two out-of-phase peaks in the outermost arrays becomes unstable nearly immediately after bifurcation from linear edge state, then it stabilizes and remains stable in the largest part of the gap, and becomes unstable when it enters the second band [modes 3,4 in Fig. 3(d)]. Finally, edge soliton with three in-phase peaks surprisingly remains stable in the entire gap [modes 5,6 in Fig. 3(d)]. Interestingly, increasing power results in gradual concentration of light in central array in this state, see mode 6.

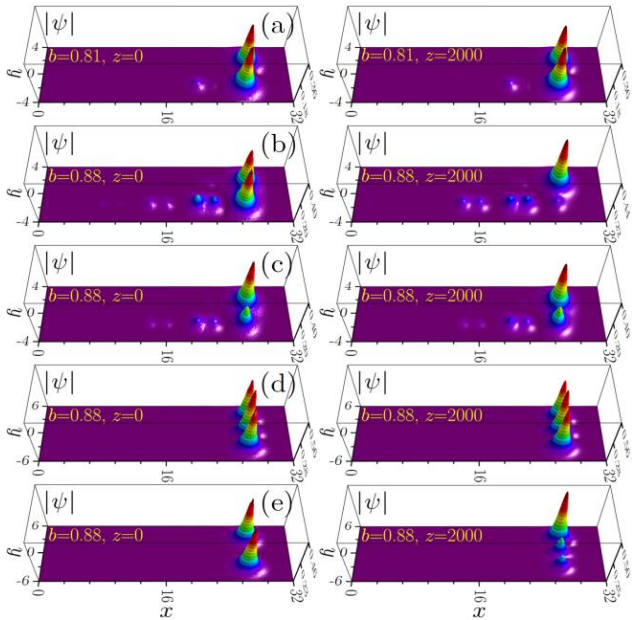

Fig. 4. Examples of stable propagation (a),(c),(d) and decay (b),(e) of edge solitons in two- (a),(c) and three- (d),(e) coupled SSH arrays. Field modulus distributions at different distances are shown. Inputs correspond to modes 1 (a), 4 (b), 5 (c) from Fig. 3(b) and modes 1 (d) and 3 (e) from Fig. 3(d).

Typical propagation scenarios of perturbed edge solitons are illustrated in Fig. 4(a)-(c) for $K=2$ and in Fig. 4(d),(e) for $K=3$ arrays. Thus, dipole [Fig. 4(a)] and asymmetric [Fig. 4(c)] solitons remain stable in $K=2$ array and maintain their internal structure over large distances. Unstable symmetric soliton [Fig. 4(b)] decays via progressively increasing oscillations of two main maxima. The example of stable evolution of soliton with three in-phase peaks in $K=3$ array is shown in Fig. 4(d). Its low-power dipole counterpart decays and transforms into asymmetric state, see Fig. 4(e).

Summarizing, we have shown that coupling between SSH arrays substantially enriches the spectrum of available topological states. The number of such states is equal to the number of coupled arrays in topological phase. They give rise to multipole edge solitons with different internal symmetries, most of which can be stable.

**Funding:** Research project FFUU-2024-0003 of the Institute of Spectroscopy of RAS.

**Disclosures:** The authors declare no conflicts of interest.

**Data availability.** Data underlying the results presented in this paper may be obtained from the authors upon reasonable request.